# Long-term electrical conductivity of titanium-implanted alumina

R.E. Spirin [a*], M.C. Salvadori [a], L.G. Sgubin [a], I.G. Brown [b]

[a] *Institute of Physics, University of São Paulo, C.P. 66318, CEP 05314-970, São Paulo S.P., Brazil*

[b] *Lawrence Berkeley National Laboratory, Berkeley, CA 94720, USA*

**Abstract**

Metal ion implantation into ceramics has been demonstrated to be an effective and controllable technique for tailoring the surface electrical conductivity of the ceramic piece, and this approach has been used in a number of applications. Importantly, it provides a method for grading the voltage drop across high voltage insulators, and thereby increasing the maximum operational voltage that can be applied across the insulator without surface flashover. However a concern for the use of the method is the long term stability of the implantation-induced conductivity, this especially so if the implanted metal species is readily oxidized. Here we report on our examination of the long-term behavior of the surface conductivity of titanium-implanted alumina. The results indicate that after an initial drop of as much as 40% in the first few weeks after implantation, the conductivity shows only a very slow decrease of about 10% over the following year. This degree of change should be quite sufficiently small to be unimportant for all but the most demanding situations.

---

[*] Corresponding author. *E-mail address*: romulsre@usp.br

## 1. Introduction

The voltage hold-off capability of high-voltage ceramic insulators in vacuum is poor compared to that of the bulk ceramic material or the vacuum itself. Electron processes can lead to a discharge across the ceramic surface, commonly referred to as surface flashover [1]. The hold-off voltage can be increased by providing some surface electrical conductivity so as to uniformly grade the voltage drop along the length of the insulator. By providing for a small, uniform leakage current, electron charge build-up is prevented. A method for providing the required surface conductivity is energetic metal ion implantation into the ceramic. This approach has been used with considerable success to increase the maximum acceptable operating voltage of accelerator columns [2,3] and to assure electrostatic field symmetry in ion microprobe applications [4]. The method has been explored by a number of workers [5–8]; metal ion species used have included Na, Al, Ti, Cu, Pt, Au and Pb, and ceramics have included alumina and glass-ceramics. The physical process by which surface electrical conductivity is induced has been explained [9–11] by a model in which an insulator-conductor composite of metal nanoparticles is formed in the implanted near-surface layer, and conductivity follows by a percolation process when the conducting nanoparticles are in geometric contact ("percolation" refers to the flow of current through random resistor networks [12]).

In many applications it is important that the implantation-induced surface conductivity not diminish significantly over a period of many years, and in this context the concern arises of the possible slow oxidation of the implanted metal. Thus Pt has been used [3,4] as the implantation species as an approach to avoid this possibility. However since Pt is an expensive material there is certainly motivation to use a lower cost metal such as Ti. Titanium also functions very nicely as a vacuum arc cathode material, which is important since most high dose metal ion implantation is done by means of a vacuum arc ion source-based implantation facility. Thus we arrive at the necessity to determine how the surface conductivity of ceramics implanted with Ti behaves in the long term, say over a period of a year or more.

We have carried out Ti ion implantation into alumina coupons so as to modify (increase) the surface electrical conductivity, and monitored the conductivity as a function of time for a period of 383 days. Here we report on this work and the results obtained.

## 2. Experimental

The implantation procedure and detailed analysis of the nanocomposite formed have been described in detail elsewhere [13], including Ti depth profiles as measured by Rutherford backscattering (RBS), conductivity measurements, TEM images of Ti nanoparticles formed in the near-surface nanocomposite layer, and comparison of the conductivity measurements with theory. In brief, we used a novel and compact vacuum arc based "inverted implanter" facility [14] to implant small alumina samples with Ti ions at 52 keV to a range of doses from about $2 \times 10^{15}$ to $6 \times 10^{16}$ ions/cm$^2$. In the inverted implanter technique, the ion source and plasma are held at ground potential while the final grid of the ion beam formation electrode system and the substrate (implantation target) are held at high negative potential, thus avoiding the experimental complication and expense associated with holding the plasma source and its electronics at high potential. The ion extraction voltage used was 25 kV. The titanium plasma (formed by a vacuum arc discharge) from which the ion beam is formed is composed of multiply charged ion species, and for titanium the charge state distribution is known [15] to be Ti$^+$:Ti$^{2+}$:Ti$^{3+}$ = 11:75:14 (particle percent), corresponding to a mean charge state $Q = 2.1$. Thus the mean ion implantation energy, eQV, was 52 keV. The implantation is done in a repetitively pulsed mode [14,16] with 100 μs pulses at a repetition rate of 3 pulses per second. A magnetically suppressed Faraday cup at the target location monitors the beam current, which was about 11 mA/cm$^2$ peak pulsed, or a time-averaged ion implantation current density of about 3.3 μA/cm$^2$.

The alumina samples were 26.2 mm long by 7.1 mm wide, and we estimate the thickness of the implanted layer to be about 55 nm [13]. As the implantation proceeded, resistance measurements were performed in situ. After a certain number of pulses of known dose (measured by the Faraday cup), the implantation process was temporarily suspended and the sample resistance measured, after which the implantation proceeded, and so on, generating data of resistance as a function of dose. The conductivity for each implantation dose was obtained considering the dimensions of the region where the nanocomposite was formed in the substrate. Beyond the detailed work described in Ref. [13], we continued to monitor the resistance of a selected implanted sample for a long period of time, during which the sample was left exposed to the normal laboratory atmosphere. The resistance was measured, and from that the conductivity calculated, with three different voltages applied across the sample: 9 V, 100 V and 1 kV; (the application of higher voltage can "bind" nanoparticles which creates more inter-particle connections and thus increases the conductivity).

**3. Results and discussion**

The measured surface resistivity, also called sheet resistance, is shown in Fig.1 as a function of implantation dose. As expected, the resistivity falls with increasing dose, spanning a range of about six orders of magnitude for the dose range explored here of $(0.2 – 6) \times 10^{16}$ cm$^{-2}$. Thus this method of surface resistivity tailoring provides the capability of targeting specific resistivity values.

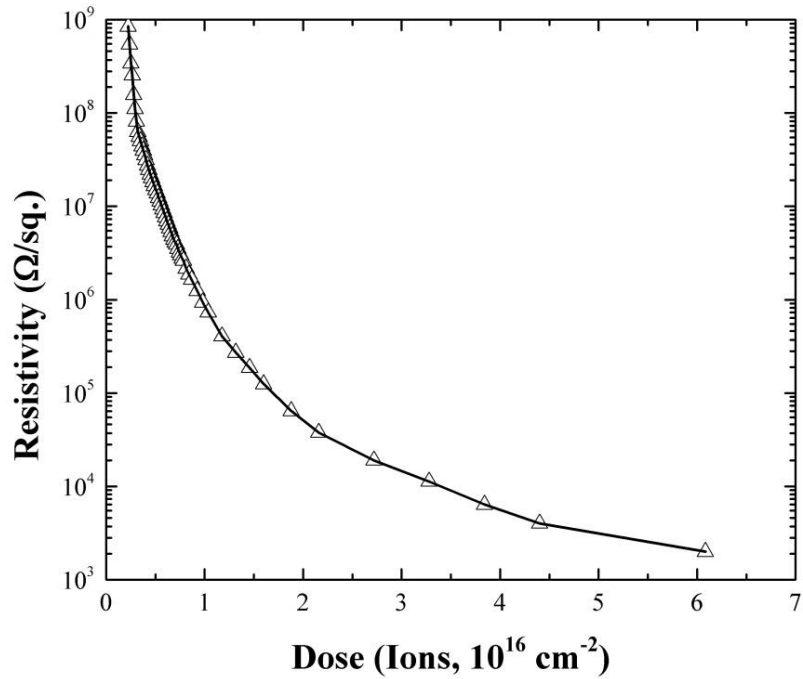

**Fig. 1.** Measured surface resistivity as a function of Ti implantation dose.

Following implantation, a selected sample was stored in the normal laboratory environment and its surface resistivity monitored as a function of time for a period of 383 days. The results are shown in Fig. 2. Here the data are shown in the form of electrical conductivity, taking the thickness of the implanted, conducting layer to be 55 nm [13]. We see that there is a drop in conductivity of a few tens of percent over the first few weeks, followed by a long period (a year) over which the conductivity drops by only about a further 10%.

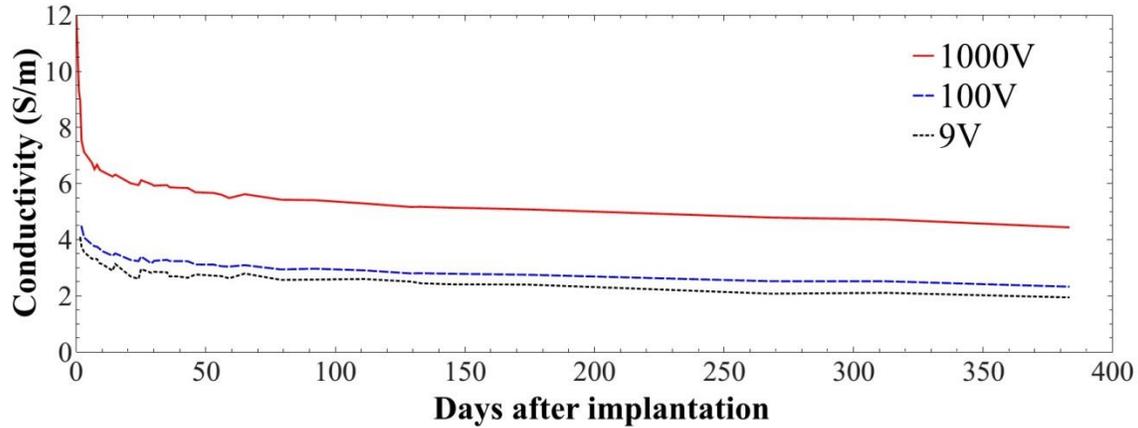

**Fig. 2.** Conductivity of Ti-implanted alumina as a function of time after implantation, with measurement voltages of 9 V, 100 V and 1000 V. The sample was exposed to the normal laboratory environment.

**4. Conclusion**

We have shown that ion implantation of Ti into alumina provides a powerful tool for the tailoring of surface resistivity. Specific values of resistivity can be targeted over a range of at least six orders of magnitude by control of the implantation dose. Importantly, the surface resistivity does not change substantially over the long term; the drop in conductivity remains quite sufficiently small to be unimportant for all but the most demanding applications. Ti can be used as the metal ion implantation species; noble metals such as Pt are not essential. We point out that typically the implanted surface will not be exposed to atmosphere, as here, but will be maintained at high vacuum for much of the time, and in this case the slow oxidation of implanted Ti and consequent decrease of conductivity over time would be expected to be much less still than measured here.

**Acknowledgments**

This work was supported by the Fundação de Amparo a Pesquisa do Estado de São Paulo (FAPESP) № 2013/10224-3 and the Conselho Nacional de Desenvolvimento Científico e Tecnológico (CNPq), Brazil.